\documentclass[twocolumn,showpacs,groupedaddress]{revtex4}
\usepackage{graphicx}

\begin{document}

\draft
\title{Hahn echo and criticality in spin-chain systems}
\author{X. X. Yi, H.  Wang, and W. Wang}
\affiliation{Department of Physics, Dalian University of
Technology, Dalian 116024, China}

\date{\today}

\begin{abstract}
We establish a relation between Hahn spin-echo of a spin-$\frac 1
2 $ particle and quantum phase transitions in a spin-chain, which
couples to the particle. The Hahn echo is calculated and discussed
at zero as well as at finite temperatures. On the example of XY
model, we show that the critical points of the chain are marked by
the extremal values in the Hahn echo, and can influence the Hahn
echo in finite  temperatures. An explanation for the relation
between the echo and criticality  is also presented.
\end{abstract}

 \maketitle

Quantum phase transitions\cite{sachdev99}(QPTs) have attracted
enormous attention within various fields of physics in the past
decade. They exist on all length scales, from microscopic to
macroscopic. Because QPTs, which describe transitions between
quantitatively distinct phases, are driven solely by quantum
fluctuations, they provide valuable information about the ground
state and nearby excited states of quantum many-body systems. The
observation of quantum criticality depends eventually on the
experimentally available temperature, then it is natural to ask
how high in temperature can the effects of quantum criticality
persist? Do quantum critical points shed light on quantum
mechanics of macroscopic systems, for instance providing a deeper
understanding of decoherence? In this paper we answer these
questions by examining exact solutions of the XY spin-chain model,
calculating the Hahn echo   of a spin-$1/2$ particle coupled to
the critical  spin-chain.

The Hahn echo was first introduced by Hahn\cite{hahn50} to observe
and measure directly transverse relaxation time $T_2$, i.e., the
dephasing time. It differentiates  from the  Loschmidt echo in
that the latter measures the sensitivity of quantum system
dynamics to perturbations in the Hamiltonian. For a certain regime
of parameters, the Loschmidt echo decays exponentially with a rate
given by the Lyapunov exponent of the underlying classically
chaotic system. Recently, a huge interest was attracted in the
attempt of characterizing QPTs in terms of entanglement, by
analyzing extremal points, scaling and asymptotic behavior in
various entanglement
measures\cite{osterloh02,vidal03,wu04,chen04,gu05}. The relation
between Berry's phases and quantum critical points was also
established recently in the XY model\cite{carollo05,zhu05,
hamma06}. In this paper, we shall show how critical points can be
reflected in the Hahn spin-echo, and what is the finite
temperature effect on the Hahn spin-echo.

Consider a spin-$1/2$ particle coupled to a spin-chain described
by the one-dimensional XY model, the Hamiltonian of such a system
may be given by
\begin{equation}
H=H_s+H_c+H_i, \label{h1}
\end{equation}
where
\begin{eqnarray}
H_s&=&\mu s^z,\nonumber\\
H_c&=&-2\sum_{l=1}^N((1+\gamma)s_l^x s_{l+1}^x+ (1-\gamma)s_l^y
s_{l+1}^y+\lambda s_l^z),\nonumber\\
H_i&=&4g\sum_{l=1}^N s^z s_l^z.\label{h2}
\end{eqnarray}
Here $s$ denotes  spin operator of the system particle which
couples to the chain spins $s_l$ $(l=1,...,N)$ located at the
lattice site $l$. The spins in the chain are coupled to the system
particle through a constant $g$. The Hahn echo experiments
consists in preparing the system spin in the initial state
$|y_s\rangle=(|\uparrow\rangle +i|\downarrow\rangle)/\sqrt{2}$,
and then allowing free evolution for time $\tau$. A $\pi$-pulse
described by the Pauli operator $\sigma^x$ is then applied to the
system spin, and after free evolution for one more interval $\tau$
an echo is observed, which provides a direct measurement of single
spin coherence. We would like to notice that the free evolution
here means no additional driving fields exist, the  coupling
between the system and the spin-chain is always there.

\begin{figure}
\includegraphics*[width=0.8\columnwidth,
height=0.6\columnwidth]{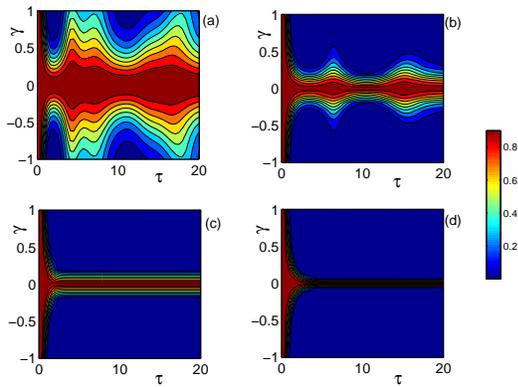} \caption{(color online)
Hahn echo of the spin-$\frac 1 2 $ particle {\it vs.} time $\tau$
and the anisotropy parameter $\gamma$. The spin-$\frac 1 2 $
particle was coupled to a spin-chain described by the XY model.
The parameters chosen are $N=246$ sites, $g=0.3$ and
(a)$\lambda=2$, (b) $\lambda=1.5$,  (c)$\lambda=1$, and
(d)$\lambda=0.5$.  } \label{fig1}
\end{figure}

\begin{figure}
\includegraphics*[width=0.8\columnwidth,
height=0.6\columnwidth]{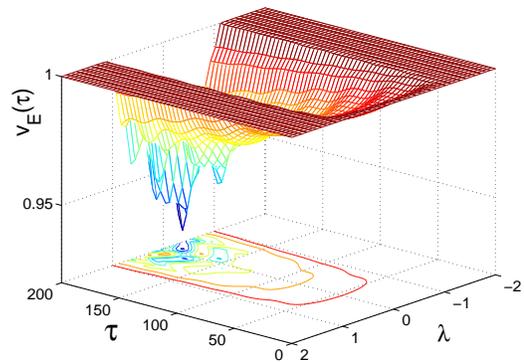} \caption{(color
online)Hahn echo as a function of time $\tau$ and $\lambda$. The
figure was plotted for $N=246$ sites, $g=0.1$ and $\gamma=0.001$.}
\label{fig2}
\end{figure}

We now follow the calculation\cite{witzel05} to derive an exact
expression for the Hahn echo decay due to the system-chain
couplings in Eq.(\ref{h1}). The density matrix for the whole
system (system and the spin-chain) which will be used to calculate
the Hahn echo is given by
\begin{equation}
\rho(\tau)=U(\tau)\rho_0 U^{\dagger}(\tau),
\end{equation}
where $U(\tau)$ denotes the evolution operator\cite{witzel05}
\begin{equation}
U(\tau)=u(\tau)\sigma^x u(\tau), u(\tau)=e^{-iH\tau},
\end{equation}
and $\rho_0$ is taken to be the initial state of the whole system
\begin{equation}
\rho_0=|y_s\rangle\langle y_s|\otimes \rho_c(0)
\end{equation}
with $\rho_c(0)$ denoting the initial state for the spin-chain.
The Hahn spin echo envelope is then given by
\begin{equation}
v_E(\tau)=2|{\mbox Tr}\{(s^x+is^y)\rho(\tau)\}|. \label{he1}
\end{equation}
In order to get an explicit expression for the Hahn echo envelope
Eq.(\ref{he1}), we first write $u(\tau)$ in basis
$\{|\uparrow\rangle, |\downarrow\rangle\}$ (the eigenstates of
$\sigma^z$), by noting that $[H_s, H_i]=0.$ This leads to
\begin{equation}
u(\tau)=\sum_{j=\uparrow,\downarrow}u_j(\tau)|j\rangle\langle j|,
\end{equation}
with $u_j(\tau)$ satisfying,
\begin{eqnarray}
i\hbar\frac{\partial}{\partial t}u_j(t)=H_j
u_j(t),\nonumber\\
H_j=-2\sum_{l=1}^N((1+\gamma)s_l^x s_{l+1}^x+ (1-\gamma)s_l^y
s_{l+1}^y+\lambda_j s_l^z),
\end{eqnarray}
where $\lambda_j=\lambda\pm g$,  $+$ and $-$ correspond to
$\downarrow$ and $\uparrow$, respectively. The free energy $\mu$
of the system which contributes only energy shifts to $H_j$ would
not affect the Hahn echo and has been omitted hereafter. For the
system initially in state $|j\rangle$ $(j=\uparrow, \downarrow)$,
the dynamics and statistical properties of the spin-chain would be
govern by $H_j$, it takes the same form as $H_c$ but with
perturbed field strengths $\lambda_j$. This perturbation to the
spin-chain regardless of how small it is  can be reflected in the
Loschmidt spin echo decay \cite{quan05}, in particular at critical
points. What behind the decay is the orthogonalization  between
two ground states obtained for two different values of external
parameters\cite{zanardi05}. The Hamiltonian $H_j$ can be
diagonalized by a standard procedure to be
\begin{equation}
H_j=\sum_k\omega_{j,k}(\eta_{j,k}^{\dagger}\eta_{j,k}-\frac 1 2 ),
\end{equation}
which can be summarized in the following three steps. (1) The
Wigner-Jordan transformation, which converts the spin operators
into fermionic operators via the relation
$a_l=(\prod_{m<l}\sigma_m^z)(\sigma_l^x+i\sigma_l^y)/2$, where
$\sigma_l$ is the Pauli matrix of the spin at site $l$; (2)The
Fourier transformation, $d_k=\frac{1}{\sqrt{N}}\sum_la_l
\mbox{exp}(-i2\pi lk/N);$ And (3) the Bogoliubov transformation,
which defines the fermionic operators,
\begin{equation}
\eta_{j,k}=d_k\cos\frac{\theta_{j,k}}{2}
-id_{-k}^{\dagger}\sin\frac{\theta_{j,k}}{2},
\end{equation}
where the mixing angle $\theta_{j,k}$ was defined by
$\cos\theta_{j,k}=\varepsilon_{j,k}/\omega_{j,k},$ with
$\omega_{j,k}=\sqrt{\varepsilon_{j,k}^2+\gamma^2\sin^2\frac{2\pi
k}{N}}, $ and  $\varepsilon_{j,k}=(\cos\frac{2\pi
k}{N}-\lambda_{j}), $ $ k=-N/2, -N/2+1,...,N/2-1.$ To diagonalize
the spin chain Hamiltonian, the periodic boundary condition was
used in this paper. The boundary term
$H_b=-(a^{\dagger}_1a_N+a_N^{\dagger}a_1+\gamma(a_N^{\dagger}a_1^{\dagger}
+h.c.))(P+1)$ with $P=exp(i\pi\sum_{j=1}^N a_j^{\dagger}a_j)$
would vanish when $N/2$ is odd. Since the paper aims at finding
the link between the Hahn echo and the critical points, we will
chose $N/2$ odd to simplify the boundary effects. This treatment
is available in the limit $N\rightarrow \infty$, where the
boundary effects are negligible.
 It is easy to show that
$[\eta_{i,k},\eta_{j,k}]\neq 0$ when $j\neq i$, i.e., the modes
$\eta_{i,k}$ and $\eta_{j,k}$ do not commute(this is not the case
for some special parameters discussed later on). This would result
in the Hahn echo decay as you will see. With these results, the
evolution operator $U(\tau)$ can be reduced to
\begin{equation}
U(\tau)=u_{\uparrow}(\tau)u_{\downarrow} (\tau)|\uparrow\rangle
\langle\downarrow|+u_{\downarrow}(\tau)u_{\uparrow}(\tau)
|\downarrow\rangle\langle \uparrow|,
\end{equation}
with
$u_j(\tau)=e^{-i\sum_k\omega_{j,k}(\eta_{j,k}^{\dagger}\eta_{j,k}-\frac
1 2 )\tau}\equiv \prod_k u_{j,k}(\tau)$, $j=\uparrow,\downarrow ,$
and $ u_{j,k}(\tau)=e^{-i
\omega_{j,k}(\eta_{j,k}^{\dagger}\eta_{j,k}-\frac 1 2 )\tau}.$
After a simple algebra, we arrive at
\begin{equation}
v_E(\tau)=|{\mbox
Tr_c}(u^{\dagger}_{\downarrow}u^{\dagger}_{\uparrow}
u_{\downarrow}u_{\uparrow}\rho_c(0))|, \label{he2}
\end{equation}
where the trace is taken over the spin-chain. Eq.(\ref{he2}) can
be simplified by noting that
($n_{j,k}=\eta_{j,k}^{\dagger}\eta_{j,k}$)
\begin{equation}
[n_{\uparrow,k},n_{\downarrow,k}]=\frac i 2
 \sin(\theta_{\uparrow,k}-\theta_{\downarrow,k})(\eta_{-k}\eta_{k}-
 \eta_{k}^{\dagger}\eta_{-k}^{\dagger}),
\end{equation}
and consequently,
\begin{equation}
u^{\dagger}_{\uparrow,k}u_{\downarrow,k}=u_{\downarrow,k}
u^{\dagger}_{\uparrow,k} +\hat{X}_k, \label{cor}
\end{equation}
where $\hat{X}_k=(1-e^{i\omega_{\uparrow,k}t})
(1-e^{-i\omega_{\downarrow,k}t})[n_{\uparrow,k},n_{\downarrow,k}]$.
Here $\eta_{k}=d_k\cos\frac{\theta_{k}}{2}
-id_{-k}^{\dagger}\sin\frac{\theta_{k}}{2},$ and
$\theta_k=\theta_{j,k}|_{\lambda_j=\lambda}.$ Substituting
Eq.(\ref{cor}) into $v_E(\tau)$, we get
\begin{equation}
v_E(\tau)=\prod_k|1+{\mbox Tr_c }[u^{\dagger}_{\downarrow,k}
\hat{X}_ku_{\uparrow,k}\rho_c(0)]|.\label{se2}
\end{equation}
The explicit expression for Eq.(\ref{se2}) can be obtained by
choosing a specific initial state of the chain. We shall consider
two initial states in this paper, (1) $\rho_c(0)$ is taken to be
the ground state of $H_c$, (2) $\rho_c(0)$ is chosen to be a
thermal state for the spin-chain. The ground state of $H_c$
follows by the same steps summarized above. It is defined as the
state to be annihilated by each operator $\eta_{k}$, namely
$\eta_k|g(\gamma,\lambda)\rangle=0.$ After a few manipulations we
obtain the Hahn echo envelope at zero temperature,
\begin{widetext}
\begin{eqnarray}
v_E(t)&=&\prod_k|1+\frac 1 4
\sin(\theta_{\uparrow,k}-\theta_{\downarrow,k})
\sin(\theta_{k}-\theta_{\uparrow,k})|1-e^{i\omega_{\uparrow,k}t}|^2(1-
e^{-i\omega_{\downarrow,k}t}-\sin^2\frac{\theta_k-\theta_{\downarrow,k}}{2}
|1-e^{i\omega_{\downarrow,k}t}|^2)\nonumber\\
&-&\frac 1 4 \sin(\theta_{\uparrow,k}-\theta_{\downarrow,k})
\sin(\theta_{k}-\theta_{\downarrow,k})|1-e^{i\omega_{\downarrow,k}t}|^2(1-
e^{i\omega_{\uparrow,k}t}-\sin^2\frac{\theta_k-\theta_{\uparrow,k}}{2}
|1-e^{i\omega_{\uparrow,k}t}|^2)|. \label{sef}
\end{eqnarray}
\end{widetext}
With the above expressions, we now turn to study the Hahn echo at
zero temperature. Since the XY model is exactly solvable and still
present a rich structure, it offers a benchmark to test the
properties of Hahn echo in the proximity of a quantum phase
transition. For the XY model one can identify the critical points
by finding the regions where the energy gap $\omega_k$ vanishes.
Indeed, there are two regions in the $\lambda$, $\gamma$ space
that are critical. Namely, $\gamma=0$ for $-1<\lambda<1$, and
$\lambda=\pm 1$ for all $\gamma$. We first focus on the
criticality in the XX model. The XX model that corresponds to
$\gamma=0$ has a criticality regime along the lines between
$\lambda=1$ and $\lambda=-1$. The critical points can be read out
from the Hahn echo as shown in figures \ref{fig1} and \ref{fig2}.
Figure 1 shows the Hahn echo as a function of time $\tau$ and the
anisotropy parameter $\gamma$. Clearly, the Hahn echo takes a
sharp change in the limit $\gamma\rightarrow 0$, this results can
be understood by considering the value of $\theta_{j,k}$ and
$\theta_k$, which take 0 or $\pi$ depending on the sign of
$\cos(2\pi k/N)-\lambda_j$ and $\cos(2\pi k/N)-\lambda$,
respectively. In either case, $\sin(\theta_k-\theta_{j,k})=
\sin(\theta_{\uparrow,k}-\theta_{\downarrow,k})=0,$ this leads to
$v_E(\tau)=1.$ Physically, when $\gamma=0$, the particle number
operators $n_{\uparrow,k}$ and $n_{\downarrow,k}$ commute, which
implies that the perturbation from the system to the spin-chain
does not excite the spin-chain, then the Hahn echo   which
characterizes the dephasing of the system remains unit. Figure 2
shows the Hahn echo $v_E(\tau)$ in the vicinity of critical points
$\gamma \rightarrow 0$ and $\lambda=\pm 1$. A sharp change among
the line of $\lambda=\pm 1$ appears clearly.

We would like to notice that the Hahn echo $v_E(\tau)$ at critical
points of $\gamma=0$ and $\lambda=\pm 1$ does not depend on the
chain-system coupling constant $g$, but in the vicinity of
$\gamma=0$, it does. This was shown in figure 3, where we plotted
the Hahn echo as a function of $\lambda$ and $g$ with
$\gamma=0.001$ (close to zero). As expected, the critical points
have been shifted linearly by the coupling constant $g$. The white
area in figure 3 corresponds to $v_E(\tau)=1$. In the region of
$g>2$ and $-1<\lambda <1$, $v_E(\tau)$ always equal to 1. This can
be understood by examining  the the definition of $\theta_{j,k}$
and $\theta_k$. In this region, $\lambda_j=\lambda\pm g\geq 1$,
leading to $\theta_{j,k}=\theta_k$ for any $k$ in the limit
$\gamma \rightarrow 0$. This results in $v_E(\tau)=1$, which is a
direct followup  of  Eq.(\ref{sef}).

Now we turn to study the criticality in the transverse Ising
model($\gamma=1$ in the XY model). The ground state structure of
this model change dramatically as the parameter $\gamma$ is
varied. We first summarize the ground states of this model in the
limits of $|\lambda|\rightarrow \infty$, $|\lambda|=1$ and
$\lambda=0$. The ground state of the spin-chain approaches a
product of spins pointing the positive/negative  $z$ direction in
the $|\lambda|\rightarrow \infty$ limit, whereas the ground state
in the limit $\lambda=0$ is doubly degenerate under the global
spin flip by $\prod_{l=1}^N\sigma_l^z$. At $|\lambda|=1$, a
fundamental transition in the ground state occurs, the symmetry
under the global spin flip breaks at this point and the chain
develops a nonzero magnetization $\langle\sigma_x\rangle\neq 0$
which increases with $\lambda$ growing.
\begin{figure}
\includegraphics*[width=0.8\columnwidth,
height=0.6\columnwidth]{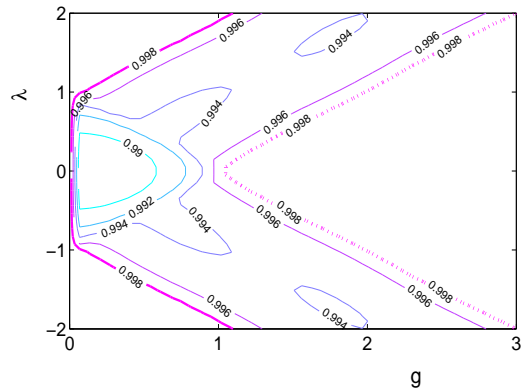} \caption{(color
online)This figure was plotted to show the dependence of the
critical points on the system-chain coupling constant $g$. Time
$\tau=50$, $N=246$ sites, and $\gamma=0.001$ were chosen for this
figure. } \label{fig3}
\end{figure}
The above mentioned properties of the ground state are reflected
in the Hahn echo as shown in figure 4. In the limit
$|\lambda|\rightarrow \infty$, $\theta_{j,k}=\theta_k=\pi/(-\pi)$,
 this results in $v_E(\tau)=1$. In fact as figure 4-(a) shows, when
$|\lambda|\geq 4$, $v_E(\tau)$ approaches 1 very well. With
$|\lambda| \rightarrow 1$, the Hahn echo $v_E(\tau)$ tends to
zero, this can be interpreted as the sensitivity of the spin-chain
ground state to perturbations from the system-chain coupling at
these points. The Hahn echo is a oscillating function of time
$\tau$ around $\lambda=0$. Due to the coupling to the spin-chain,
the oscillation is damping, and eventually $v_E(\tau)$ tends to
zero in the $\tau \rightarrow \infty$ limit.
\begin{figure}
\includegraphics*[width=0.8\columnwidth,
height=0.8\columnwidth]{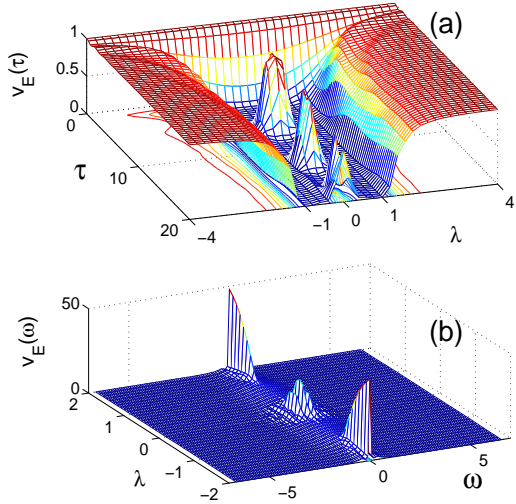} \caption{(color
online)(a)Hahn echo {\it versus} time $\tau$ and $\lambda$ with
$\gamma=1$. The other parameters chose are $g=0.1$, $N=246$ sites.
(b) Discrete Fourier transformation of $v_E(\tau)$, with the same
parameters as in figure (a). } \label{fig4}
\end{figure}
The difference between cases of $\gamma=0$ and $\gamma=1$ is that
$[n_{\uparrow,k},n_{\downarrow,k}]=0$ for $\gamma=0$, but it does
not hold for $\gamma=1$. This is the reason why the Hahn echo
takes different values at these critical points.Figure 4-(b) is a
discrete Fourier transformation of $v_E(\tau)$ with the same
parameters as in figure 4-(a). It would provides us the Hahn echo
in the frequency domain.  The ground state of the XY model is
really complicated with many different regime of
behavior\cite{barouch70}, these are reflected in sharp changes in
the Hahn echo across the line $|\lambda|=1$ regardless of $\gamma$
(as shown in figure 5, except $\gamma=0$), indicating the change
in the ground state of the spin-chain from paramagnetic phase to
the others.
\begin{figure}
\includegraphics*[width=0.8\columnwidth,
height=0.8\columnwidth]{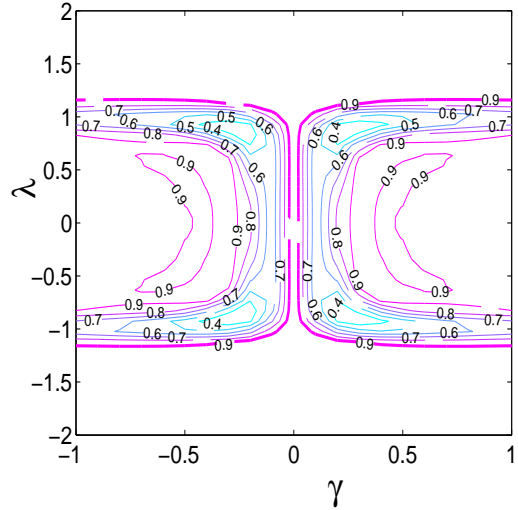} \caption{(color
online)Hahn echo at time $\tau=20$ with $N=246$ sites, and
$g=0.01$. } \label{fig5}
\end{figure}

Up to now, we did not consider the temperature effect. Finite
temperature is the regime to which all experiments being confined,
but what is the finite temperature effect on the Hahn echo? In the
following, we shall consider this problem by studying the
contributions of one- and two-particle excitations to the Hahn
echo. Taking a thermal state $\rho_c^T(0)=\frac 1 z e^{-\beta
H_c}$ $(\beta=\frac{1}{k_B T}$) as the initial state of the
spin-chain, the Hahn echo envelope can be written as
\begin{equation}
v_E^T(\tau)=\prod_k|1+\frac 1 z \sum_n e^{-\beta \Omega_n}\langle
n|u_{\downarrow,k}^{\dagger} \hat{X}_k u_{\uparrow,k} |n\rangle|,
\end{equation}
where $|n\rangle$ and $\Omega_n$ denote the eigenstate and
corresponding eigenvalue of $H_c$, respectively. $z$ is the
partition function.
\begin{figure}
\includegraphics*[width=0.9\columnwidth,
height=0.45\columnwidth]{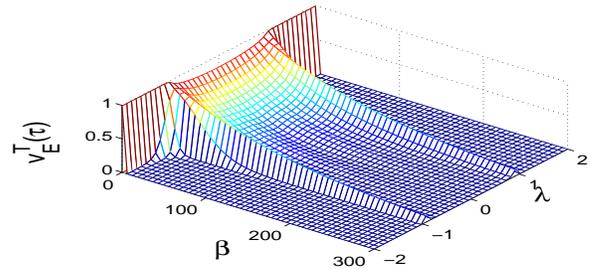} \caption{(color online)
 Contribution of the two-particle excitation to Hahn echo at time
 $\tau=40$. The other parameters chosen are $N=246$ sites, $\gamma=0.01$,
 and $g=0.02$.} \label{fig6}
\end{figure}
We shall restrict our consideration to the contribution from one-
and two-particle excitations of the chain, namely,
\begin{eqnarray}
|n\rangle \in \{ \eta_{j,k}^{\dagger}|g(\gamma,\lambda)\rangle,
\eta_{i,k_1}^{\dagger}\eta_{j,k_2}^{\dagger}|g(\gamma,\lambda)\rangle\},
\nonumber\\ \mbox{ for } k_1\neq k_2, \mbox{\ or \ } i\neq j,
\end{eqnarray}
with $k$, $k_1$ and $k_2$ ranging from $-N/2$ to $N/2-1$. It is
not difficult to show that there are no contribution from the one
particle excitation, because $\hat{X}_k$ creates or annihilates
two particles with $k$ and $-k$ jointly. The numerical results
presented in figure 6 show the contribution of the two-particle
excitation to the Hahn echo, we find that the quantum critical
points can influence the Hahn echo at a finite temperature. For
the parameters chosen in figure 6, the contribution form the
thermal excitation is larger than that from quantum fluctuation
when $\beta <72=\beta_c$. Here we have scaled out an overall
energy scale denoted by $J$. $J$ may be taken to be of order
$1000K$, that is the order for the antiferromagnetic exchange
constant of the Heisenberg model. It yields $T_c \sim 14K$
corresponding to parameters chosen in figure 6. For the transverse
Ising model $\gamma=1$, $\beta_c$ is of order $10$, we obtain
$T_c\sim 100K$ in this situation with the other parameters being
the same as in figure 6. Notice that the study here is based on
the Hahn echo(a dynamical quantity), this would differ from the
investigation based on thermodynamics\cite{kopp05}. We would like
to notice that the discussion on the finite temperature effect was
limited to very low temperatures, because only one- and
two-quasiparticle excitations were included. Nevertheless, it is
interesting because it also sheds light on the contribution to
Hahn echo from the first excited states, which have the same
energy as the ground state $\rho_c(0)$ of $H_c$ at critical
points. The results presented in figure 6 show that those
contributions tend to zero with $T\rightarrow 0$.

In conclusion, by discussing the Hahn spin echo in the spin-$\frac
1 2 $ particle coupled to  critical spin-chains, the relation
between the Hahn echo and the critical points was established. The
relation not only provides an efficient theoretical tool to study
quantum phase transitions, but also proposes a method to measure
the critical points in experiments.  Up to two-particle
excitations, we have also studied the influence of thermal
fluctuation on the Hahn echo, it would shed light on the low
temperature(with respect to the overall energy scale $J$) effects
on the Hahn echo.

\ \ \\
This work was supported
by NCET of M.O.E, and NSF of China Project No. 10305002 and 60578014.\\

\end{document}